\documentclass[11pt]{article}
\usepackage{amssymb,epsf,graphicx}
\usepackage{amsmath}

\allowdisplaybreaks[1]

\newcommand{\sectiono}[1]{\section{#1}\setcounter{equation}{0}}

\begin{document}
{}~ \hfill\vbox{\hbox{LMU-ASC 17/08}}\break
\vskip 2.1cm

\centerline{\Large \bf A tachyon lump in closed string field theory}

\vspace*{8.0ex}

\centerline{\Large \rm Nicolas Moeller}

\vspace*{3.0ex}

\centerline{\large \it Arnold-Sommerfeld-Center for Theoretical Physics}
\centerline{\large \it Department f\"ur Physik, Ludwig-Maximilians-Universit\"at M\"unchen}
\centerline{\large \it Theresienstra\ss e 37, 80333 M\"unchen, Germany} 
\vspace*{2.0ex}
\centerline{E-mail: {\tt nicolas.moeller@physik.uni-muenchen.de}}

\vspace*{6.0ex}

\centerline{\bf Abstract}
\bigskip

We find a codimension one lump solution of closed bosonic string field
theory. We consider vertices up to quartic order and include in the
string field the tachyon, the ghost dilaton, and a metric fluctuation.
While the tachyon profile clearly is that of a lump, we observe that
the ghost dilaton is roughly constant in the direction transverse to
the lump, equal to the value it takes in the nonperturbative tachyon
vacuum. We explain, with a simple model, why this should be expected.

\vfill \eject

\baselineskip=16pt

\tableofcontents


\sectiono{Introduction}
\label{s_intro}

There is new evidence coming from closed string field theory (CSFT)
\cite{CSFT}, that there is a nonperturbative closed bosonic string
vacuum.  This tachyon vacuum was first found\footnote{A
  nonperturbative vacuum was already found in \cite{Kos-Sam} on the
  CSFT truncated to cubic order. To this order, however, the ghost
  dilaton doesn't participate, and interesting features of the vacuum
  thus do not appear.}  by Yang and Zwiebach \cite{vacuum} by using
covariant closed string field theory truncated to quartic order
\cite{quartic}. The calculation was subsequently improved by including
more levels of massive fields \cite{Moe-Yang}, and the quintic vertex
\cite{quintic}. A computation of the effective potential of the
tachyon and ghost dilaton by integrating out all massive fields
\cite{quintic2} shows that the properties of the tachyon vacuum do not
change much when the quintic term is included; this is evidence that
truncating the CSFT action to a finite order is a meaningful
approximation. The effective potential also shows that the tachyon
vacuum is a saddle point; we will shortly comment later why this
doesn't necessarily mean that it is unstable.

The nature of the tachyon vacuum remains mysterious, but it has been
conjectured that it corresponds to the state of the universe after a
big crunch (\cite{Yang:2005rw}, \cite{vacuum}). To test this idea
further, it is interesting to construct dynamical or solitonic
solutions; this is partially motivated by evidence from $p$-adic
string theory that indicates that closed tachyon lumps could be
noncritical string theories \cite{Moe-Sch}. In \cite{Berg-Raza},
Bergman and Razamat found that it is possible to find such lump
solutions in the low-energy effective action of the tachyon, dilaton
and metric. They found that the dilaton must grow far away from the
lump, and that it has a constant gradient along it; this is a nice
check that the lump corresponds to a noncritical string theory with
linear dilaton background.

In this paper we use covariant closed string field theory to construct
a lump depending on one spatial coordinate. While one may argue that
two spatial coordinates are needed in order to construct a lump with a
linear dilaton background along it, interestingly enough we do find a
lump solution with one codimension. Actually, in the last part of
their paper \cite{Berg-Raza}, Bergman and Razamat already attempted to
find a codimension one lump solution in CSFT. They found a lump whose
field configuration sits in the tachyon vacuum far from the core of
the lump, and reaches and overshoots the perturbative vacuum (i.e.
vanishing tachyon and dilaton) on the core. We make a few criticisms
on their calculation: They didn't include the metric perturbation
which, as we will see later, must be included when we have
non-homogeneous string fields; they considered only second derivatives
of the tachyon field and neglected all the higher derivatives; and
they also neglected the kinetic terms of the dilaton and massive
fields. While the first two problems can be considered an
approximation, the third one seems to be in contradiction with their
result (their dilaton profile has an amplitude of the same order of
magnitude as that of the the tachyon profile, as can be seen on their
Figure 5 (a)), and therefore unjustified. As we will see, the dilaton
kinetic term, and more particularly its sign, will play an important
role in our calculation. It is perhaps interesting to note that its
magnitude on our solution is, however, rather small; it should become
clear at the end of this section, that the dilaton must in fact avoid
gaining kinetic energy.

\begin{figure}[!ht]
\begin{center}
\input{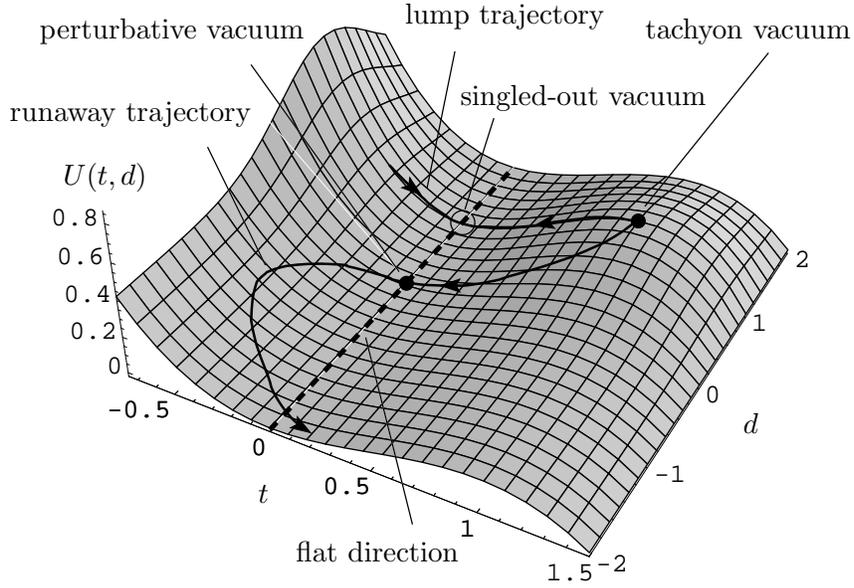}
\caption{\footnotesize{Two trajectories on the pseudo-potential: One
    that goes through the perturbative vacuum but then runs away, and
    a lump solution that comes back to its starting point, but does
    not go through the $(t,d) = (0,0)$ perturbative vacuum, instead
    it ``selects'' another perturbative vacuum.}}
\label{toypotential_f}
\end{center}
\end{figure}

\paragraph{}
It is instructive to consider a very simple (but still related to
CSFT) toy model of one tachyon $t$ (in this paper $t$ denotes the
tachyon, time will not appear in the equations) and one dilaton $d$
given by the action
\begin{equation}
S = \int d^D x \left( -\frac{1}{2} \partial_\mu t \, \partial^\mu t + \frac{1}{2} 
\partial_\mu d \, \partial^\mu d - V(t,d) \right).
\end{equation}
Our convention for the metric is $\eta_{\mu \nu} =
\text{diag}(-1,1,\ldots,1)$. This is not much different from the
calculation of Bergman and Razamat, except that we have been careful
to take the right sign in front of the dilaton's kinetic term. The
potential $V(t,d)$ has a local maximum at $(t,d) = (0,0)$ where it is
zero, a flat direction along the dilaton axis, and a saddle point at,
say $(t_0,d_0) = (1,1)$ where it is negative. Now let us suppose that
the fields $t$ and $d$ depend on one spatial coordinate $x$. Let us
start to look at the tachyon only, which has a regular kinetic term.
Its equation of motion is
\begin{equation}
t''(x) = \frac{\partial V}{\partial t}.
\end{equation}
This is simply the very well known fact that a solution $t(x)$ can be
seen as the time-dependent trajectory of $t$ in the inverted potential
$-V$. Now let us look at the dilaton only, which has a kinetic term of
the opposite sign. The equation of motion for $d(x)$ is
\begin{equation}
d''(x) = -\frac{\partial V}{\partial d}.
\end{equation}
In other words, the profile $d(x)$ can be seen as the trajectory of
$d$ in the potential itself. Now let us consider both fields together,
evolving in the potential $V(t,d)$. The equations of motion are
\begin{equation}
t''(x) = \frac{\partial V(t,d)}{\partial t}
\ , \qquad
d''(x) = -\frac{\partial V(t,d)}{\partial d},
\end{equation}
and the question is: can we understand the solution of this system of
equations as the trajectory of $(t,d)$ in some ``pseudo-potential''? In other
words, can we find $U(t,d)$ such that
\begin{equation}
t''(x) = -\frac{\partial U(t,d)}{\partial t}
\ , \qquad
d''(x) = -\frac{\partial U(t,d)}{\partial d} \quad ?
\end{equation}
We can immediately answer this question negatively by noting that it implies
\begin{equation}
  \frac{\partial U(t,d)}{\partial t} = -\frac{\partial
    V(t,d)}{\partial t}
\qquad \text{and} \qquad 
  \frac{\partial U(t,d)}{\partial d} = \frac{\partial
    V(t,d)}{\partial d},
\end{equation}
and considering the multiple derivative $\partial_t \partial_d
U(t,d)$. From the first equation we obtain $\partial_t \partial_d
U(t,d) = - \partial_t \partial_d V(t,d)$, while the second equation
tells us that $\partial_t \partial_d U(t,d) = \partial_t \partial_d
V(t,d)$. These two equations are clearly incompatible unless
$\partial_t \partial_d V(t,d) = 0$. We conclude from this simple
example, that it is hard to reach an intuitive understanding of the
lump profile when the two fields' kinetic terms have different signs.
But we nevertheless attempt a guess. In our situation, we know that
the potential $V(t,d)$ at the origin, has a local maximum in the
tachyon direction and is flat along the dilaton direction. In this
special case where $\partial_d V(t,d)=0$, we can obviously view the
dynamics as happening in the inverted potential $-V(t,d)$. Next, we
know that the potential has a saddle point at $(t_0, d_0) = (1,1)$. We
will assume that (maybe after some field redefinition) the tachyon
corresponds to the ``minimum direction'', whereas the dilaton
corresponds to the ``maximum direction''\footnote{It is interesting to
  observe that, due to the unusual sign of the dilaton's kinetic term,
  the saddle point is actually {\em stable}!}. In terms of these
redefined (and normalized) fields, the potential can be written
locally as
\begin{equation}
  V = \frac{1}{2} (t-1)^2 - \frac{1}{2} (d-1)^2 + \ 
  \text{terms of cubic order.}
\end{equation}
In particular we have $\partial_t
\partial_d V(t,d) = 0 + \text{linear terms}$. And, to leading order,
the equations of motion can be understood as the motion in the
pseudo-potential
\begin{equation}
U = - \frac{1}{2} (t-1)^2 - \frac{1}{2} (d-1)^2,
\end{equation}
a local maximum at $(1, 1)$. We will thus have a flat valley along the
$t=0$ axis and a local maximum at $(1,1)$. We don't make any
assumption about the rest of the pseudo-potential as it might not be
defined globally. A function $U$ with the above properties up to
numerical factors, is given by
\begin{equation} 
  U(t,d) = \left(1 +
    e^{-(d-1)^2} \right) \left( \frac{1}{2} t^2 - \frac{1}{3} t^3
  \right).  
\end{equation} 
On Figure \ref{toypotential_f} we show two trajectories
on this pseudo-potential.
It is clear that a solution that starts from the tachyon vacuum $(t,d)
= (1,1)$ (here a local maximum) and goes through the perturbative
vacuum $(t,d) = (0,0)$, will continue forever towards the negative
dilaton direction. A lump solution, however, should come back to the
tachyon vacuum. Such a solution will necessarily pass through the
point $(t,d) = (0,1)$. In contrast with the result of
\cite{Berg-Raza}, the lump does not go through the perturbative vacuum
$(0,0)$. This should not be surprising because the potential has a
flat direction and we have, in fact, a family of perturbative vacua
along this flat direction. By choosing $(t,d)=(0,0)$ as perturbative
vacuum, we are merely arbitrarily choosing one of them. It is of
course well known that different vacua along the flat dilaton
direction, are related by a change in the closed string coupling
constant. After having made this basic observation, it should be clear
that {\em there is no reason why the lump should go through the point
  $(t,d)=(0,0)$, it could go through any other perturbative vacuum
  along the flat dilaton direction}\footnote{In the order- and
  level-truncated CSFT, the dilaton direction is only approximately
  flat and therefore only finitely many vacua exist nearby. Since
  $(t,d) = (0,0)$ is a real extremum of the approximate potential, one
  may argue that we should still expect the lump to pass by this
  point. But the lump spends only a finite amount of ``time''
  (``distance'' would be more precise) near the flat direction, it is
  therefore not too sensible to the imperfections of our
  approximation; so, as we will see, the lump in fact does not go near
  the point $(0,0)$.}. From our simple model and Figure
\ref{toypotential_f}, we may rather suggest that {\em the ghost
  dilaton will remain roughly constant}. In this paper, we will show
that this is indeed what is happening in CSFT. Finally we take a
slightly different point of view: Not knowing the tachyon vacuum, we
would be unable to favor one perturbative vacuum from another along
the flat dilaton direction (or in other words, a coupling constant
from another). But now, the lump solution actually singles out one
preferred perturbative vacuum, the one it crosses on its core. In
other words, {\em the closed tachyon lump solution singles out one
  value of the coupling constant}. We will not speculate further along
this line.

\paragraph{}
In the next section, we construct the codimension one lump solution in
closed string field theory, and in the last section we will discuss
its relation to the low-energy effective action.

\sectiono{Tachyon lump in closed string field theory}
\label{s_lump}

We are considering the covariant closed string field theory action
\cite{CSFT} truncated to quartic order \cite{quartic}. And we use the
convention $\alpha'=2$.
\begin{equation}
S = -\frac{1}{\kappa^2} \left( \frac{1}{2} \langle
\Psi | c_0^- Q_B |
\Psi \rangle + \frac{1}{3!} \left\{ \Psi, \Psi, \Psi \right\} +
\frac{1}{4!} \left\{ \Psi, \Psi, \Psi, \Psi \right\} + \ldots \right),
\end{equation}
where $c_0^\pm = \frac{1}{2} (c_0 \pm \bar{c}_0)$ and $Q_B$ is the
BRST charge. We define the level of a state as the eigenvalue of $L_0
+ \bar{L}_0 + 2$. The momentum thus contributes to the level; this is
the obvious generalization of the level in open SFT which was shown in
\cite{lumps} to give rise to a well-convergent truncation scheme for
states with momentum. The string field to massless level in the Siegel
gauge is
\begin{equation}
  |\Psi\rangle = \int \frac{d^{26}p}{(2\pi)^{26}}
  \left( t(p) \, |T;p\rangle - \frac{1}{2}h_{\mu \nu}(p) \, |H^{\mu\nu};p\rangle 
    + d(p) \, |D;p\rangle \right),
\label{psiint}
\end{equation}
where
\begin{equation}
  |T;p\rangle = c_1 \bar{c}_1 |0;p\rangle \ , \quad
  |H^{\mu\nu};p\rangle = \alpha_{-1}^\mu\bar{\alpha}_{-1}^\nu c_1 \bar{c}_1 |0;p\rangle
  \ , \quad \text{and} \quad
  |D;p\rangle = (c_1 c_{-1} - \bar{c}_1 \bar{c}_{-1}) |0;p\rangle
\end{equation}
are respectively the tachyon, a metric fluctuation, and the ghost
dilaton with momentum $p$.  The quadratic term is easily calculated.
With the convention
\begin{equation}
\langle p| c_{-1}\bar{c}_{-1}c_0^-c_0^+c_1 \bar{c}_1 |p'\rangle = (2
\pi)^{26} \delta^{(26)}(p-p')
\end{equation}
we find
\begin{align}
  \kappa^2 S^{(2)} & = -\frac{1}{2} \langle \Psi| c_0^- Q_B
  |\Psi\rangle
  \nonumber \\
  & = \int \frac{d^{26}p}{(2\pi)^{26}} \left( \left(1 - \frac{p^2}{2} \right)
    \, t(p) t(-p) - \frac{p^2}{4} \, h_{\mu \nu}(p) h^{\mu \nu}(-p) + p^2
    \, d(p) d(-p) \right).
\label{quadratic}
\end{align}
It is important to note that, as discussed in the introduction, the
kinetic term of the dilaton has the unusual sign. In the sigma-model
as well, the dilaton comes with the irregular sign; however, after
expressing the action in terms of the Einstein metric, the sign of the
kinetic term of the dilaton becomes regular.

The cubic terms can already be a little bit complicated (see
\cite{Kos-Sam} for a complete expression), so we introduce from now on
the assumption that only one spatial component, say the $I$ component,
of the momenta is nonzero (or equivalently that the solution depends
only on the coordinate $X^I$).  It follows that we need to keep only
the component $h_{II}$ of the metric fluctuation.  Indeed the other
components can be consistently set to zero because the terms $\left\{
  h_{\mu\nu}, h_{\mu I}: \mu, \nu \neq I \right\}$ always appear at
least in pair in the action. We can thus write $h_{II}(p) = h(p)$,
which is the trace of the metric fluctuation, the {\em matter
  dilaton}. The cubic term is then (with $K = 3 \sqrt{3}/4$)
\begin{align}
  \kappa^2 S^{(3)} & = -\frac{1}{6} \left\{ \Psi, \Psi, \Psi \right\}
  = -\frac{1}{6} \int \frac{d^{26}p_1}{(2\pi)^{26}} 
  \frac{d^{26}p_2}{(2\pi)^{26}} K^{- p_1^2-p_2^2-(p_1+p_2)^2} \Bigl( 
\nonumber \\
& 2 \, K^6 \, t(p_1) t(p_2) t(-p_1-p_2) 
- \frac{3}{4} \, K^4 \, (p_1 - p_2)^2 
\, t(p_1) t(p_2) h(-p_1-p_2)
\nonumber \\
& + \frac{3}{32} \, K^2 \, \left( 4 - 
(p_1 + 2 p_2) (2 p_1 + p_2) \right)^2 
\, h(p_1) h(p_2) t(-p_1-p_2) 
\nonumber \\
& - \frac{1}{32} \, \left( (p_1 + 2 p_2) 
(2 p_1 + p_2) (p_1 - p_2) \right)^2 
\, h(p_1) h(p_2) h(-p_1-p_2)
\nonumber \\
& - 3\, K^2 \, d(p_1) d(p_2) t(-p_1-p_2)
+ \frac{3}{8} \, (p_1 - p_2)^2 
\, d(p_1) d(p_2) h(-p_1-p_2)
\Bigr).
\label{cubic} 
\end{align} 

\subsection{Lone tachyon at cubic order}
\label{s_cubic}

\begin{figure}[!ht]
\begin{center}
\input{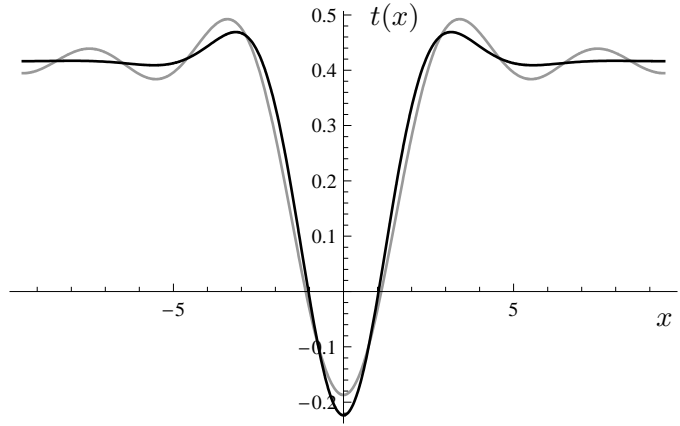}
\caption{\footnotesize{The pure cubic tachyon lump with $R=3$, 
and $N_t=4$ (gray curve) and $N_t=10$ (black curve).}}
\label{cubiclump_f}
\end{center}
\end{figure}

We start by truncating the CSFT action to cubic order. Since, to this
order, the ghost dilaton always appears quadratically into the action,
we can consistently ignore it. The matter dilaton $h$, however, can
couple linearly to tachyons with nonzero momenta; we will thus first
consider only the tachyon, then we will include the matter dilaton in
the next subsection. From (\ref{quadratic}) and (\ref{cubic}), we
write the cubic action for the tachyon
\begin{equation}
\kappa^2 S = \frac{1}{2} \, \int \frac{d^{26}p}{(2\pi)^{26}} (2 - p^2)
    \, t(p) t(-p)
- \frac{1}{3} \, \int \frac{d^{26}p_1}{(2\pi)^{26}} 
  \frac{d^{26}p_2}{(2\pi)^{26}} 
K^{6-p_1^2-p_2^2-(p_1+p_2)^2} t(p_1) t(p_2) t(-p_1-p_2),
\end{equation}
The equation of motion then reads, after Fourier transforming back to
position space
\begin{equation}
(2+\partial_x^2) K^{-2 \partial_x^2} \, \tilde{t}(x) = K^6 \, \tilde{t}(x)^2
\qquad \text{where} \qquad \tilde{t}(x) \equiv K^{\partial_x^2} \, t(x).
\label{cubicequ}
\end{equation}
At cubic order, the tachyon vacuum is at 
\begin{equation}
t_0 = 2 K^{-6} \approx 0.41620.
\end{equation}
In order to solve (\ref{cubicequ}) for a lump, we now compactify the
direction $x = X^I$ on a circle of radius $2 \pi R$, and expand the
solution $t(x)$ in an even Fourier series
\begin{equation}
t(x) = \sum_{n=0}^{N_t} t_n \, \cos(n x/R).
\end{equation}
Plugging this into Equ. (\ref{cubicequ}), one obtains a system of
nonlinear equations for the coefficients $t_n$ that can be solved
numerically. The numerical solution $t(x)$ will be meaningful if it
converges in both limits $N_t \rightarrow \infty$ and $R \rightarrow
\infty$. We do find a lump solution with these properties.  We
consider $R=3$ and $N_t = 4$ (which is a consistent bound with respect
to level truncation since here the highest tachyon mode $t_4$ has
level $16/9$. Had we included $t_5$, which has level $25/9$, we should
have also included the massless fields of level two in order to have a
consistent level truncation). We find
\begin{equation}
t(x) = 0.33106 - 0.16679 \, \cos(x/3) -0.15448 \, \cos(2 x/3) 
-0.12403 \, \cos(3 x/3) - 0.07283 \, \cos(4 x/3), \label{tachsol}
\end{equation}
which is plotted with a gray line on Figure \ref{cubiclump_f}. It is
interesting to ignore level truncation consistency and add further
harmonics to the tachyon. With $N_t = 10$ we find
\begin{align}
t(x) &= 0.335385 - 0.159796 \, \cos(x/3) -0.151211 \, \cos(2 x/3)
-0.125671 \, \cos(3x/3) 
\nonumber \\
& - 0.077683 \, \cos(4 x/3) - 0.032177 \, \cos(5 x/3) -0.009685 \, 
\cos(6 x/3) -0.002342 \, \cos(7 x/3) 
\nonumber \\
& -0.000467 \, \cos(8 x/3) -0.000076 \, \cos(9 x/3) -0.000010 \, \cos(10 x/3),
\end{align}
which is shown as a black curve on Figure \ref{cubiclump_f}.
It is interesting to observe that the tachyon profile oscillates with
a damped amplitude as we move away from the core of the lump. This
behavior is typical for a solution of an equation of the form
(\ref{cubicequ}) with a differential operator of the form
$(2+\partial_x^2) K^{-2 \partial_x^2}$; it was already observed in
\cite{Aref} in the context of tachyon kinks in cubic super SFT.

\paragraph{}
Now we would like to estimate the tension of this soliton. It is given by
\begin{equation}
{\cal T} = 2 \pi R \left(V(\text{lump}) - V(\text{vacuum}) \right),
\end{equation}
where $V(\text{vacuum})$ is the value of the potential evaluated at
the nonperturbative tachyon vacuum.  For a time-independent solution,
the potential is minus the action, so (truncating to cubic order)
\begin{equation}
V = \frac{1}{\kappa^2} \left( \frac{1}{2} \langle
\Psi | c_0^- Q_B |
\Psi \rangle + \frac{1}{3!} \left\{ \Psi, \Psi, \Psi \right\} \right).
\end{equation}
With the solution (\ref{tachsol}) we find
\begin{equation}
{\cal T} = 0.23285 \, \kappa^{-2}, \label{Ttach}
\end{equation}
and we have checked that, to a good approximation, it doesn't depend
on the compactification radius $R$. We have no physical interpretation
for this number, but it is at least a good check that the lump
solution is not a pure gauge transformation of the tachyon vacuum.

\subsection{Including the metric fluctuation}
\label{s_metric}

\begin{figure}[!ht]
\begin{center}
\input{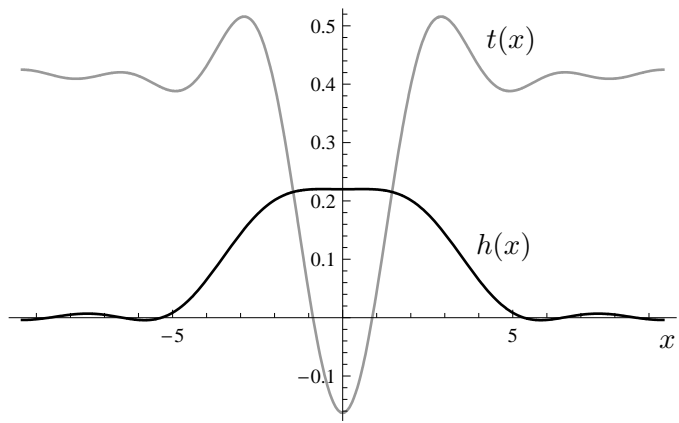}
\caption{\footnotesize{Tachyon (gray line) and matter dilaton (black line) profiles, 
with $R=3$, $N_t = 5$ and $N_h=4$.}}
\label{tachgravlump_f}
\end{center}
\end{figure}

The interesting thing now is to consider the effect of the lump on the
gravity sector, namely the matter and ghost dilatons. In the next
step, we stay at cubic order but go to level $L$ with $2 \leq L < 4$,
so that we must include the massless fields, but do not need to
include massive fields. Since the cubic vertex can couple only an even
number of ghost dilatons, we consistently set this field to zero for
now.  Instead of solving directly the equations of motion, as we did
in the last section, we will write the action in terms of the tachyon
and matter dilaton modes, and then find numerically an extremum of the
action.  The approaches are strictly equivalent and they lead to
exactly the same numerical solutions, but the latter is a bit simpler
when the equations of motion become complicated. Namely, we write
\begin{equation}
|\Psi\rangle = \sum_{n=0}^{N_t} t_n \, c_1 \bar{c}_1 \,
\frac{1}{2}(|0;n/R\rangle + |0;-n/R\rangle) 
- \frac{1}{2} \sum_{n=0}^{N_h} h_n \, \alpha_{-1}^I 
\bar{\alpha}_{-1}^I c_1 \bar{c}_1 \,
\frac{1}{2}(|0;n/R\rangle + |0;-n/R\rangle).
\end{equation}
This is the same as (\ref{psiint}) with 
\begin{align}
t(p) &= (2 \pi)^{26} \sum_{n=0}^{N_t} t_n \, \frac{1}{2}\left(\delta(p-n/R) + 
\delta(p+n/R)\right) \label{tn} \\
h(p) &= (2 \pi)^{26} \sum_{n=0}^{N_h} h_n \, \frac{1}{2}\left(\delta(p-n/R) + 
\delta(p+n/R)\right) \label{hn},
\end{align}
and with the normalization changed accordingly to the compact case
\begin{equation}
  \langle 0; n/R| c_{-1} \bar{c}_{-1} c_0^- c_0^+ c_1 \bar{c}_1 |0; m/R \rangle = \delta_{n+m}.
\end{equation}
So we can substitute (\ref{tn}) and (\ref{hn}) into (\ref{quadratic})
and (\ref{cubic}) in order to write the action in terms of the modes
$t_n$ and $h_n$. We take again $R=3$, and for the level of the fields
to be lower than four (so we don't need to include massive fields), we
must have $N_t^2/R^2 < 4$ and $N_h^2 / R^2 < 2$, we will thus take
$N_t = 5$ and $N_h=4$.  We find the following extremum of the action
corresponding to a lump solution:
\begin{align}
t_0 &= 0.35297 & t_1 &= - 0.12624 & t_2 &= -0.12959 & 
t_3 &= -0.12792 & t_4 &= -0.09243 & t_5 &= -0.03974
\nonumber \\
h_0 &= 0.07976 & h_1 &= 0.11970 & h_2 &= 0.04397 & 
h_3 &= -0.00769 & h_4 &= -0.01571
\end{align}
The profiles $t(x) = \sum_{n=0}^{N_t} t_n \, \cos(n x/R)$ and 
$h(x) = \sum_{n=0}^{N_h} h_n \, \cos(n x/R)$ are plotted on Figure 
\ref{tachgravlump_f}.
We observe that the matter dilaton takes a roughly constant value of
about $h \approx 0.22$ in the core of the lump, and zero far away from
the lump. For the tension of the lump, we find
\begin{equation}
{\cal T} = 0.19166 \, \kappa^{-2},
\end{equation}
not too different from (\ref{Ttach}).

\subsection{Going to quartic order and including the ghost dilaton}
\label{s_dilaton}

\begin{figure}[!ht]
\begin{center}
\input{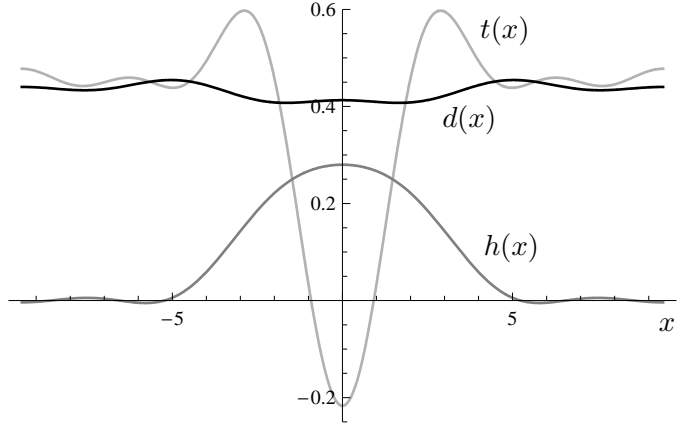}
\caption{\footnotesize{Tachyon (light gray line), matter dilaton (dark gray
    line) and ghost dilaton (black line) profiles, with $R=3$, $N_t = 5$ and
    $N_h=4$.}}
\label{tachgravdillump_f}
\end{center}
\end{figure}

We now include the quartic terms. For each term, we will need to
perform an integration over the reduced moduli space of the quartic
contact term ${\cal V} _{0,4}$. Details can be found in
\cite{dilaton}, \cite{vacuum}, \cite{quartic}, and \cite{Moe-Yang}.
Schematically, we have
\begin{equation}
\left\{ \Psi_1, \Psi_2, \Psi_3, \Psi_4 \right\} = -\frac{2}{\pi} \, 
\int_{{\cal V}_{0,4}} dx dy \, \langle \Psi_1, \Psi_2, \Psi_3, \Psi_4 
\rangle_\xi,
\end{equation}
where $\langle \Psi_1, \Psi_2, \Psi_3, \Psi_4 \rangle_\xi$ is
calculated by 1) mapping the vertex operators $\Psi_i$ from their
local coordinates to a uniformizer on the sphere, 2) insert the
antighost giving the measure on the moduli space, and 3) calculate the
resulting correlator on the sphere $\Sigma_\xi$ with punctures at $0$,
$1$, $\infty$ and $\xi = x + y \, i$. For example, for four tachyons
$|\Psi_i\rangle = c_1 \bar{c}_1 |0; p_i \rangle$, we have
\begin{equation}
  \langle \Psi_1, \Psi_2, \Psi_3, \Psi_4 \rangle_\xi = 
  (2 \pi)^{26} \delta^{(26)}\left(\sum_{i=1}^4 p_i\right) |\xi|^{2 p_1 \cdot p_3}
  |1-\xi|^{2 p_2 \cdot p_3} \prod_{i=1}^4 \rho_i(\xi, \bar{\xi})^{-2 + p_i^2},
\end{equation}
where $\rho_i$ are the mapping radii; their numerical expressions can
be found in \cite{quartic}.  The important point is that the momentum
dependence is highly nontrivial, it is not a simple function like
$K^{-p_i^2}$ that we have in the cubic vertex; but it is an integral
over reduced moduli space, parameterized by $\{p_1, p_2, p_3, p_4 \}$.
It would, in particular, be very hard to write an equation of motion
as we did in Section \ref{s_cubic}. Fortunately, in our setup of level
truncation and compact coordinate $X^I$, we need only compute a finite
number of terms. Our string field is now
\begin{align}
|\Psi\rangle & = \sum_{n=0}^{N_t} t_n \, c_1 \bar{c}_1 \,
\frac{1}{2}(|0;n/R\rangle + |0;-n/R\rangle) 
- \frac{1}{2} \sum_{n=0}^{N_h} h_n \, \alpha_{-1}^I 
\bar{\alpha}_{-1}^I c_1 \bar{c}_1 \,
\frac{1}{2}(|0;n/R\rangle + |0;-n/R\rangle)
\nonumber \\
& + \sum_{n=0}^{N_d} d_n \, (c_1 c_{-1} - \bar{c}_1 \bar{c}_{-1}) \,
\frac{1}{2}(|0;n/R\rangle + |0;-n/R\rangle).
\end{align}
At quartic order without massive fields, the tachyon vacuum is at 
\begin{equation}
(t_0, d_0) = (0.45933, 0.43878).
\end{equation}
We take again $R=3$. And we take $N_t=5$ and $N_h = N_d = 4$, which
are the maximal values allowed by consistency of level truncation.
(With this values, the action contains 14 quadratic terms, 96 cubic
terms, and 1183 quartic terms.) We find again an extremum of the action
corresponding to a lump
\begin{align}
t_0 &= 0.39264 & t_1 &= -0.13780 & t_2 &= -0.15585 & 
t_3 &= -0.15875 & t_4 &= -0.10640 & t_5 &= -0.05067 \nonumber \\
h_0 &= 0.08881 & h_1 &= 0.13897 & h_2 &= 0.06093 & 
h_3 &= 0.00281 & h_4 &= -0.01164  \nonumber \\
d_0 &= 0.43151 & d_1 &= -0.01523 & d_2 &= -0.01343 & 
d_3 &= 0.00156 & d_4 &= 0.00841 
\end{align}
The corresponding profiles are shown on Figure \ref{tachgravdillump_f}.
We note that the ghost dilaton is roughly constant along the lump, in
agreement with the simplified model discussed in the introduction.  To
estimate the tension of the lump, we include the quartic term in the
potential
\begin{equation}
V = \frac{1}{\kappa^2} \left( \frac{1}{2} \langle
\Psi | c_0^- Q_B |
\Psi \rangle + \frac{1}{3!} \left\{ \Psi, \Psi, \Psi \right\} 
+ \frac{1}{4!} \left\{ \Psi, \Psi, \Psi \Psi \right\} \right).
\end{equation}
And we find
\begin{equation}
{\cal T} = 0.25506 \, \kappa^{-2},
\end{equation}
again not too far from the value (\ref{Ttach}).

\sectiono{Discussion}
\label{s_discussion}

In \cite{Berg-Raza}, Bergman and Razamat studied closed tachyon lumps in 
the low-energy effective theory of the tachyon, dilaton and graviton 
\begin{equation}
  S = \frac{1}{2\kappa^2}\int d^Dx \sqrt{-g} e^{-2 \Phi} \left(
    R + 4 (\partial_{\mu}\Phi)^2 - (\partial_{\mu}T)^2 - 2 V(T) \right)
\end{equation}
This action was first considered in \cite{Yang:2005rw}; its equations
of motion are
\begin{align}
R_{\mu\nu} + 2 \nabla_\mu \nabla_\nu \Phi 
- (\partial_\mu T) (\partial_\nu T) &= 0 \label{Ricci} \\
\nabla^2 T - 2 (\partial_{\mu} \Phi) (\partial^{\mu}T) - V'(T) & = 0 \\
\nabla^2 \Phi - 2 (\partial_{\mu} \Phi)^2 - V(T) &= 0
\end{align}
We shortly summarize the calculation of \cite{Berg-Raza}.
After making the ansatz $T = T(x^1)$, $\Phi = \Phi(x^1, x^2)$ and 
\begin{equation}
ds^2 = (dx^1)^2 + a(x^1)^2 \eta_{\alpha \beta} dx^{\alpha} 
dx^{\beta}, \qquad \alpha, \beta = 0,2,3,\ldots D-1, \label{BRansatz}
\end{equation}
the equations of motion imply that the general form of $\Phi(x^1,
x^2)$ must be
\begin{equation}
\Phi(x^1, x^2) = {\cal D}(x^1) + Q x^2 \label{Phi}
\end{equation}
for some constant $Q$ and some function ${\cal D}$. And they also imply the constraint
\begin{equation}
Q a' = 0. \label{choice}
\end{equation}
Bergman and Razamat considered $a' = 0$ and set $a=1$, and thus found the 
equations of motion
\begin{align}
2 {\cal D}'' - (T')^2 &= 0 \label{BR1} \\
T'' - 2 {\cal D}' T' - V'(T) &= 0 \\
{\cal D}'' - 2 ({\cal D}')^2 - 2 Q^2 - V(T) &= 0 \label{BR3}
\end{align}
They found $Q \neq 0$, a linear dilaton along the $x^2$ direction. And
interestingly, from a qualitative analysis of the equations of motion
they could argue that ${\cal D}(x^1)$ must grow in both directions
$x^1 \rightarrow \pm \infty$ away from the core of the lump. They
concluded that this is consistent with a noncritical string theory in
$D-1$ dimensions in a linear dilaton background. Looking back at
(\ref{choice}), we can now understand that our lump corresponds to the
other choice, namely $Q=0$. Indeed we took from the beginning $\Phi =
\Phi(x^1)$ and included the component $h_{11}$ of the metric, so that
$a'$ need not be zero. Actually this is not quite so because our
metric is of the form
\begin{equation}
ds^2 = b(x^1)^2 (dx^1)^2 + \eta_{\alpha \beta} dx^{\alpha} 
dx^{\beta} \ , \qquad b(x^1)^2 = 1 + h(x^1), \label{bansatz}
\end{equation}
which is conformally related to (\ref{BRansatz}); but it is trivial
because in terms of
\begin{equation}
y(x^1) = \int_0^{x^1} b(\chi) d\chi,
\end{equation}
it is simply $ds^2 = (dy)^2 + \eta_{\alpha \beta} dx^{\alpha}
dx^\beta$.  In this new coordinates the equations of motion become
\begin{align}
2 \frac{d^2 \Phi}{dy^2} - \left(\frac{dT}{dy}\right)^2 &= 0 \\
\frac{d^2T}{dy^2} - 2 \frac{d\Phi}{dy} \frac{dT}{dy} - V'(T) &= 0 \\
\frac{d^2\Phi}{dy^2}  - 2 \left(\frac{d\Phi}{dy}\right)^2 - V(T) &= 0.
\label{noQ}
\end{align}
These are the same as (\ref{BR1})-(\ref{BR3}) except that we lack a
$-2 Q^2$ in the left hand side of Equ. (\ref{noQ}). It thus seems that
we have an over-constrained set of equations and that we shouldn't
find any solution; turning on one component of the metric was just
trivial. On the other hand we can argue that our equations are not
more over-constrained than (\ref{BR1})-(\ref{BR3}) once we allow a
nonzero cosmological constant because we find exactly the same
equations after replacing $V(T)$ with $V(T)+2Q^2$. The triviality of
the metric corresponds to a gauge transformation in the CSFT side.
Namely, it was shown in \cite{vacuum} that, to linear order in the
fields, the gauge-invariants are
\begin{equation}
T = t \qquad \text{and} \qquad \Phi = d + \frac{h}{4}. \label{ginvar}
\end{equation}
We can then choose a gauge where $p^\mu h_{\mu\nu} = 0$, meaning
$h=0$, and we have trivialized the metric perturbation. Equ.
(\ref{ginvar}) also tells us the relation, to linear order, between
the CSFT fields and the fields in the effective action. Looking again
at Figure \ref{tachgravdillump_f}, on can see that the linear
combination $\Phi = d + \frac{h}{4}$ will stay roughly constant along
the lump, and is not much different from $d$. An extension of this
discussion would likely require the computation of higher order terms
in the gauge transformation of the string field (see \cite{Mich} for
an interesting discussion of the relation between the fields); it is
possible that these terms will make the dilaton profile less flat.

\paragraph{}
In conclusion, the physical meaning of our lump solution is not clear
since it has no dilaton gradient along the lump, and maybe not even
across it; but at least we know that it has a nonzero tension and that
it is therefore not pure gauge. Actually, the constancy of the ghost
dilaton along the lump was expected from the very simple model
discussed in the introduction. It would be interesting to extend this
calculation by considering a ghost dilaton depending on two
codimensions $x^1$ and $x^2$ and see if it develops a gradient along
$x^2$. The main difficulty seems to be that a linear dilaton cannot be
expressed as a periodic function, so we cannot simply compactify $x^2$
as we did with $x^1$. At last, it might also be interesting to
consider lightlike solutions \cite{Hel-Sch}.

\section*{Acknowledgments}
I thank B.~Zwiebach for useful comments, and O.~Bergman and S.~Razamat
for correspondence. This work was supported in parts by the Transregio
TRR 33 'The Dark Universe' and the Excellence Cluster 'Origin and
Structure of the Universe' of the DFG.


\end{document}